\documentclass[preprint,pre,aps,superscriptaddress,a4paper]{revtex4}
\usepackage[utf8x]{inputenc}
\usepackage{graphicx, 
}
\usepackage{epstopdf}
\usepackage{amsmath}
\usepackage{amsfonts}
\usepackage{color}
\begin{document}
\title{
Combined  density functional and Brazovskii theories for systems with spontaneous inhomogeneities}
\author{ A. Ciach}
\affiliation{Institute of Physical Chemistry,
 Polish Academy of Sciences, 01-224 Warszawa, Poland}
  \date{\today} 
\begin{abstract}
 The low-T  part of the phase diagram in self-assembling systems is correctly predicted by the
 known versions of the density functional theory (DFT). The high-T part obtained in DFT, however,
 does not agree with simulations even on the qualitative level.
 In this work, a new version of the DFT  for systems with spontaneous inhomogeneities on 
 a mesoscopic length scale is developed.
 The contribution to the grand thermodynamic potential
 associated with mesoscopic fluctuations is explicitly taken into account. 
 The expression for this contribution is obtained by the methods known from the Brazovskii field theory. 
 
 Apart from developing the approximate expression for the grand thermodynamic potential
 that contains the fluctuation contribution 
 and is ready for numerical minimization, we develop a simplified version of the theory valid
 for weakly ordered phases,
 i.e. for the high -T part of the phase diagram. The simplified theory is verified by a comparison 
 with the results of 
 simulations for a particular version of the short-range attraction long-range repulsion (SALR)
 interaction potential. Except from the fact that in our theory the ordered phases are stable 
 at lower T than in simulations,
 a good agreement for the high-T part of the phase diagram is obtained for the range of density that was 
 considered in simulations. 
 In addition, the equation of state and compressibility isotherms are presented. 
 Finally, the physical interpretation of 
 the fluctuation-contribution to the grand potential is discussed in detail. 
 
\end{abstract}
\maketitle
\section{Introduction}
\begin{figure}[h]
 \centering
  \includegraphics[scale=0.26]{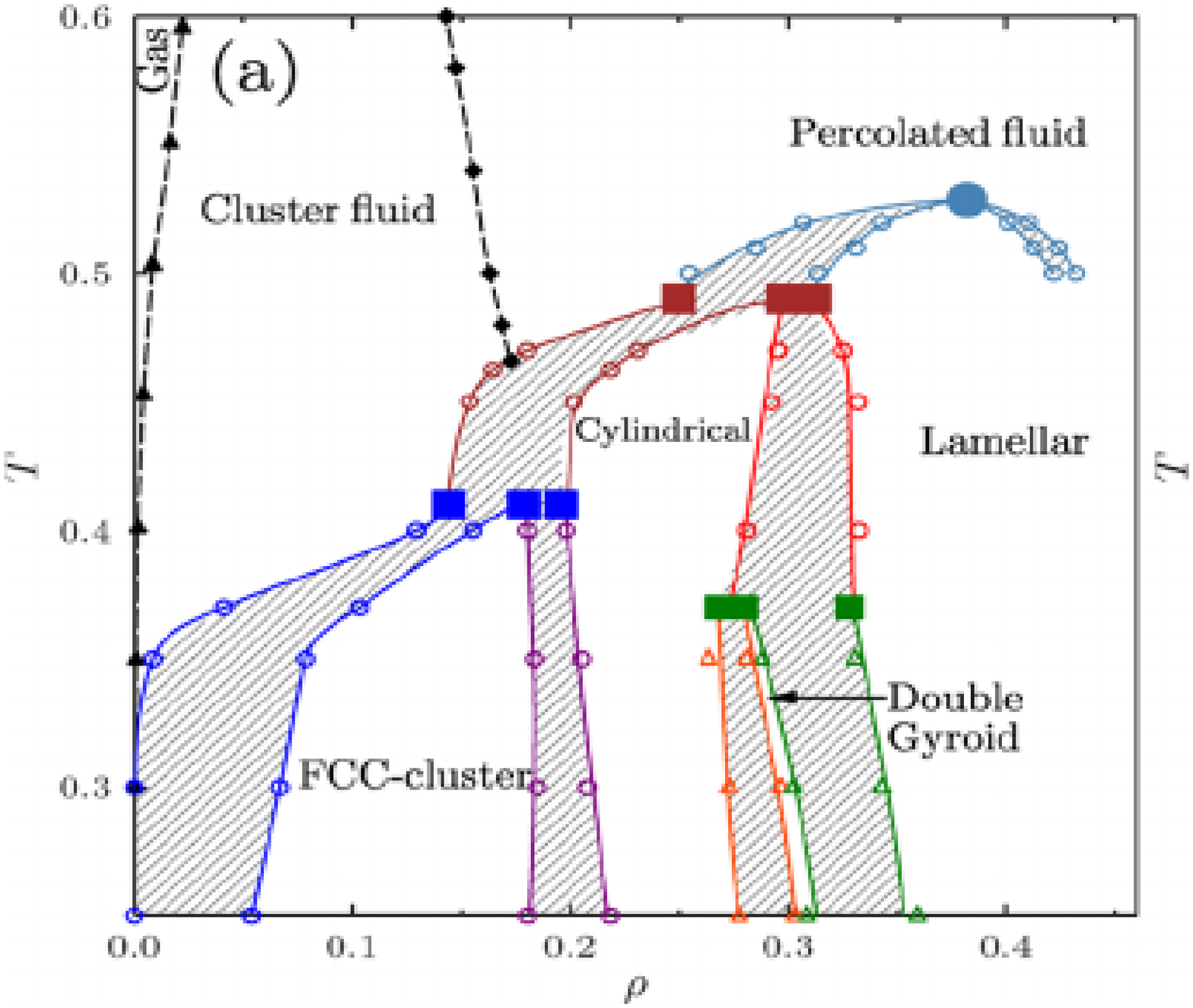}
 \includegraphics[scale=1]{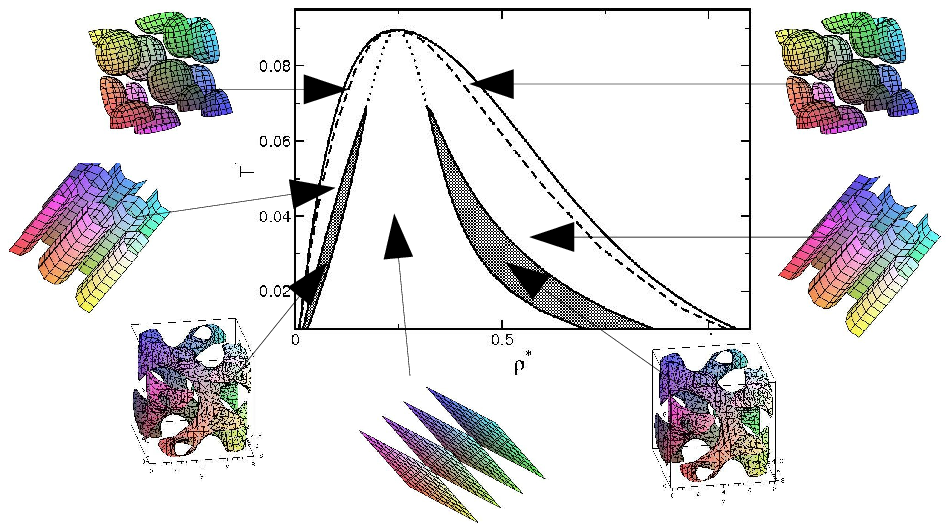}
\caption{The phase diagram in the SALR system. Panel (a):
MC simulation results for a particular form of 
the SALR potential, reprinted from Ref.\cite{zhuang:16:1}.
Panel (b): 
 phase diagram  in the MF version of the mesoscopic theory that
 is further developed in this work, reprinted from Ref.~\cite{ciach:13:0}.
Structure of the ordered phases is visualized in the cartoons surrounding the diagram.
At the shown surfaces, the local volume fraction  of the particles
is $\zeta({\bf r})=\bar\zeta$, where
 $\bar\zeta$ is the space-averaged volume fraction of the particles. 
 In the lamellar L phase the surfaces separate the alternating 
regions rich and poor in the particles. 
In the  H and C  phases stable for $\rho^{*}<0.25$, the 
volume fraction inside the cylinders and spheres  is larger than $\bar\zeta$,
while in the inverted IH and IC phases stable for $\rho^{*}>0.25$, $\zeta({\bf r})>\bar\zeta$
outside the cylinders and spheres. The dimensionless density
$N\sigma^3/V$ is 
denoted by $\rho$ and $\rho^*$ in panels (a) and (b) respectively. Temperature $T$
is in different reduced units in the two panels.  $T^*$ in panel (b) should be multiplied by about $22$ 
for comparison with panel (a).
Note that at high $T$ the phase diagrams are qualitatively different; the L phase 
in panel (a) coexists with the disordered phase, while in panel (b) it coexists with the H and IH phases.
Moreover, at high $T$ the L phase in simulations is stable at significantly
higher density ($\rho\approx 0.39$) than in MF ($\rho\approx 0.25$). 
In a more accurate DFT, a phase diagram qualitatively 
similar to the one shown in panel (b) was obtained~\cite{edelmann:16:0,pini:17:0}. In particular,
 at high $T$ the sequence of phases is
D,C,H,L,IH,IC,D, instead of D,L,D obtained in simulations. }
\label{Phases}
\end{figure}

Spontaneously appearing
aggregates such as  clusters, networks 
 or layers of particles, as well as a distribution in space of these objects, 
pose a real challenge for experiment, 
theory and simulation~\cite{royall:18:0,edelmann:16:0,pini:17:0,ciach:08:1,zhuang:16:1}. 
Recently a generic model
of self-assembly, where the particles immersed in a solvent interact with effective short-range attraction
and long-range 
repulsion (SALR), has been studied intensely by various methods 
\cite{sear:99:0,pini:00:0,imperio:04:0,pini:06:0,archer:07:1,archer:08:0,chacko:15:0,ciach:10:1,
pekalski:13:0,candia:06:0,imperio:06:0,ciach:08:1,ciach:13:0,pekalski:14:0,almarza:14:0,
edelmann:16:0,zhuang:16:0,zhuang:16:2,zhuang:16:1,pini:17:0,zhuang:17:0}. As a result,
general features of the phase diagram in the  SALR system are already 
known~\cite{imperio:06:0,ciach:08:1,ciach:13:0,pekalski:14:0,almarza:14:0,edelmann:16:0,zhuang:16:0,zhuang:16:1,pini:17:0}.
For sufficiently low temperatures,
the sequence of structures for increasing volume fraction of the particles is: disordered (D phase), 
cluster crystal with a cubic symmetry (C phase),
hexagonally ordered  cylindrical clusters (H phase), parallel  layers (L phase), hexagonally ordered cylindrical voids 
(IH phase),  cubic crystal of spherical voids (IC phase), and again the disordered  phase.
In addition, a gyroid G phase is stable between the H and L phases, and the IG phase is stable between the L and IH phases 
for some temperature range~\cite{ciach:08:1,zhuang:16:0,pini:17:0,ciach:10:1}.
Theories of mean-field nature predict the above
sequence of phases for the whole range of $T<T_L$, where above $T_L$ only the disordered phase
(no periodic structure)
is stable for the whole range of the volume fractions~\cite{ciach:08:1,archer:08:0,pini:17:0}. 
Simulations, however, show that the C phase looses stability at $T=T_C$, 
and for $T_C<T<T_H$ the disordered phase coexists with the hexagonal phase. For $T>T_H$ the hexagonal phase 
disappears, and for  $T_H<T<T_L$ the  disordered phase coexists with the lamellar 
phase~\cite{zhuang:16:0,zhuang:16:1}. Moreover, in simulations the ordered phases are stable
at higher densities and
lower temperatures than predicted by the mean-field (MF) theories. The phase diagram obtained in MF and 
in simulations, and the
structure of the ordered phases are shown in Fig.\ref{Phases}.

The periodic structure is destroyed by 
long-wavelength fluctuations (displacements or reshaping of the aggregates) that play increasingly important 
role for increasing $T$, but are neglected
in the MF theories. For this reason the MF theories cannot predict 
the phase diagram that agrees with simulations 
for relatively high temperature. 
The fluctuations are taken into account in the Landau-Brazovskii (LB) 
theory~\cite{brazovskii:75:0}, and indeed, the coexistence of 
the disordered and lamellar phases is obtained in this theory when the fluctuations are taken into 
account within the 
field-theoretic (FT) framework \cite{podneks:96:0}. Unfortunately, the LB theory is of phenomenological 
nature and the functional of 
the order parameter (OP) depends on phenomenological parameters. 
A relation of these parameters with the measurable quantities cannot be determined within the LB theory.

An attempt to combine the density functional theory (DFT) with the LB theory
has been undertaken in Ref.\cite{ciach:08:1,ciach:11:0,ciach:12:0,ciach:16:0,ciach:16:1}.
In this approach, the 
short- and long-wavelength fluctuations of the local volume fraction are included in
two separate contributions to the
grand potential. 
The first contribution has the standard DFT form in the local density approximation.
The second contribution is associated with the long-wavelength fluctuations
and has the form known from the statistical field theory. In this approach, the grand potential functional,
$\Omega[\zeta]$,  depends
on the mesoscopic volume fraction $\zeta({\bf r})$ that represents the microscopic volume fraction
averaged over the mesoscopic region (somewhat larger than the size of 
the particles, and smaller than the size of the aggregates) around ${\bf r}$. The equilibrium structure 
corresponds to the minimum of the grand-potential functional. The equations obtained in Ref.\cite{ciach:08:1,ciach:11:0},
however, are rather difficult, and  so far the phase diagram
 has been obtained in this theory only on the MF level
~\cite{ciach:08:1,ciach:10:1}. 
The effects of fluctuations have been taken into account in determination of the equation of state (EOS)
for the disordered phase~\cite{ciach:12:0,ciach:16:0,ciach:16:1}. 
It is worth mentioning that the presence of aggregates leads to a 
significant change of the shape of the lines $\mu(\bar\zeta)$ and $p(\bar\zeta)$, where $\mu$ and $p$ are the 
chemical potential and pressure respectively, and $\bar\zeta$ is the space-averaged volume fraction of the particles.
In particular, for $\bar\zeta$ optimal for a periodic structure, 
the compressibility is quite small, despite rather small value of $\bar\zeta$,
and is large for  $\bar\zeta$ that does not fit any ordered pattern.
The predictions of our theory were compared with the exact results obtained in a one-dimensional model
\cite{pekalski:13:0},
and a semiquantitative agreement was obtained~\cite{ciach:16:1}. 
In particular, at low $T$, i.e. close to the stability of the 
periodic structure at $T=0$, a step-like shape of $\mu(\bar\zeta)$ was obtained,
in agreement with the exact results. Thus, the theory is promising 
and is worth  further development.

In this work we make additional assumptions concerning the dominant fluctuation-contribution to the
grand potential. With these assumptions, we 
obtain in sec.\ref{theory} an expression for $\Omega[\zeta]$ that can be directly minimized numerically.
In sec.\ref{weak}, we
limit ourselves to relatively high $T$, where the average volume fraction $\zeta({\bf r})$ has a nearly
sinusoidal shape~\cite{pini:17:0} in the direction of oscillations. 
In this case, the average volume fraction $\zeta({\bf r})$ can be characterized
by its space-averaged value $\bar \zeta$, the period of oscillations $2\pi/k_0$, 
 the amplitude of the oscillations $\Phi$ and by 
 the symmetry of the ordered structure.
 From minimization of $\Omega$ we
 obtain equations for  $\Phi$, $\mu(\bar \zeta)$ and $p(\bar \zeta)$, and from the latter two we get $p(\mu)$ 
 for the stable or metastable structures. 
 These results allow for a construction of the high-$T$ portion of the phase diagram. 
 In sec.\ref{examples}, we consider the 
 SALR systems studied before by simulations in Ref.\cite{zhuang:16:0,zhuang:16:1}.
 We obtain the high-$T$ part of the phase diagram and compare our results with simulations.
 In addition, we obtain and discuss the EOS and compressibility isotherms. Finally, we present $\Phi(\bar\zeta)$
 in the ordered phases and compare it with the fluctuation (standard deviation of the local volume 
 fraction from $\bar\zeta$) in the metastable D phase for the corresponding thermodynamic state. 
 In sec.\ref{discussion}, we discuss the effects of spontaneously formed mesoscopic inhomogeneities in the D phase on 
 the internal energy and entropy,
 and argue that in our theory the fluctuation contributions to the grand potential have a clear physical interpretation.  
 We summarize our results in the same section. 

\section{derivation of the Grand-potential functional of the mesoscopic volume fraction of particles}
\label{theory}

\begin{figure}[h]
 \centering
  \includegraphics[scale=0.27]{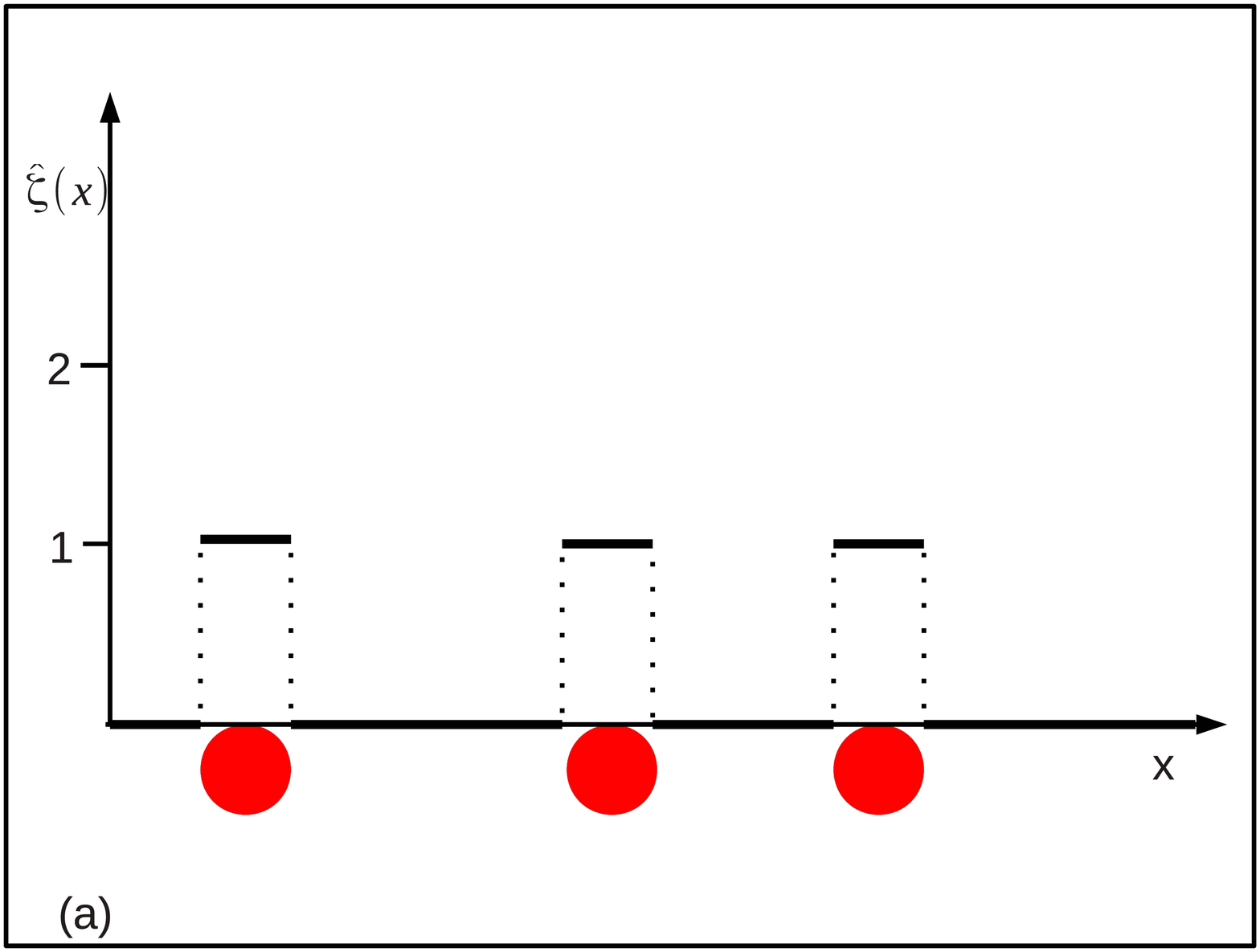}
 \includegraphics[scale=0.25]{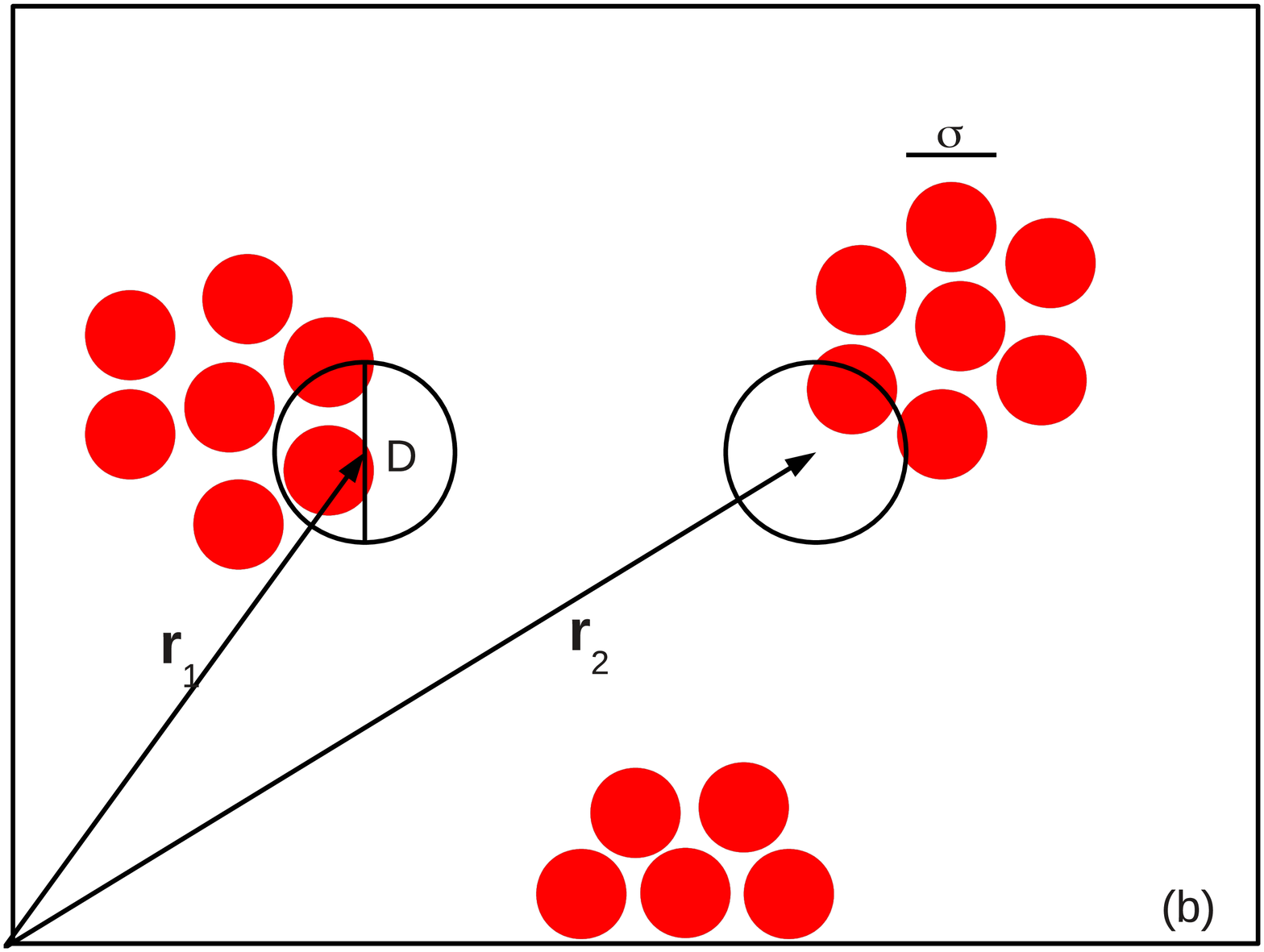}
 \caption{Panel (a): the microscopic volume fraction in the one-dimensional 
 case for the microscopic state
 represented by the red circles. Panel (b):  Construction of the mesoscopic volume fraction
 in a two dimensional system with the particles self-assembling into small clusters separated by a distance 
 larger than the range of the repulsion. The mesoscopic regions are shown as the circles
 with the diameter $D$ and the centers at ${\bf r}_1$ and ${\bf r}_2$.
 $\zeta({\bf r}_i)$ is the fraction of the area of the circle that is covered by the particles. 
 The coarse-graining procedure leads to a continuous function $\zeta({\bf r})$, 
 at the cost of smearing of the clusters. }
 \label{cartoon}
 \end{figure}

The mesoscopic volume fraction has been introduced in Ref.\cite{ciach:11:0}.
Here we briefly summarize its key properties.
Consider first the microscopic volume fraction for $N$ spherical particles with the diameter $\sigma$ 
and the centers at ${\bf r}_{\alpha}$,
 $\hat\zeta({\bf r})=\sum_{\alpha=1}^N\theta (\sigma/2-|{\bf r}-{\bf r}_{\alpha}|)$, 
 where $\theta$ is the Heaviside step function (Fig.\ref{cartoon}a).
  In the case of the macroscopic volume $V$,
  $\frac{1}{V}\int d{\bf r}\hat\zeta({\bf r})=\frac{\pi}{6}\sigma^3N/V=\bar\zeta$, where $\bar\zeta$ 
  is the macroscopic volume fraction of the particles. We can define the local mesoscopic volume fraction at 
  ${\bf r}$ in a similar way as in the formula above, by averaging over the sphere with the center at ${\bf r}$ and the diameter 
   $\sigma\le D\ll\lambda$, where $\lambda$ is the scale of the inhomogeneities in the system (Fig.\ref{cartoon}b). 
   The precise value of $D$ 
   has no significant effect on the results, as long as we are interested in the structure
   formation on the larger length scale $\lambda$.  Note that by construction,
   $\zeta({\bf r})$ is a continuous function
   and $0\le \zeta({\bf r})\le \zeta_{cp}$,
   where  $\zeta_{cp}$ is the close-packing volume fraction.
   In this mesoscopic theory, we can describe the distribution of the clusters or layers, but cannot
   describe the structure inside the aggregates. 

  $\zeta({\bf r})$ can be considered
 as a constraint imposed on the microscopic volume fractions~\cite{ciach:08:1,ciach:11:0}.  
 The constraint $\zeta({\bf r})$ means that in the allowed microstates the particles occupy
 the fraction  $\zeta({\bf r})$
 of  the mesoscopic volume  around ${\bf r}$.
 In the presence of the constraint  $\zeta({\bf r})$, the grand potential has the form~\cite{ciach:08:1,ciach:11:0}
 \begin{equation}
\label{Omco}
\Omega_{co}[\zeta]=U[\zeta]-TS[\zeta]-\mu \int d{\bf r}\zeta({\bf r}),
\end{equation}
 where $U[\zeta],S[\zeta]$ and $\mu$ are the internal energy, the entropy and the chemical potential
 respectively in the system with the constraint $\zeta({\bf r})$ imposed on the microscopic volume fractions. 
 
  When fluctuations $\phi({\bf r})$ around $\zeta({\bf r})$ can occur, 
 they lead to an extra contribution to the  grand potential, and
 \begin{eqnarray}
 \label{Ome}
  \beta \Omega[\zeta]= \beta \Omega_{co}[\zeta]-\ln\Bigg[
  \int D\phi e^{-\beta H_f[\zeta,\phi]}
  \Bigg]
 \end{eqnarray}

 where 
 
\begin{eqnarray}
\label{Hf1}
 \beta H_f[\zeta,\phi]=\beta \Omega_{co}[\zeta+\phi]-\beta \Omega_{co}[\zeta],
\end{eqnarray}
$\beta=1/k_BT$ and $k_B$ is the Boltzmann constant.
$\zeta$ represents the {\it average} mesoscopic  volume fraction when $\langle\phi\rangle=0$, and $\Omega[\zeta]$ 
takes the minimum, i.e. 
\begin{eqnarray}
\label{mincond}
 \frac{\delta \beta\Omega[\zeta]}{\delta \zeta({\bf r})}=\frac{\delta \beta\Omega_{co}[\zeta]}{\delta\zeta({\bf r})}
 +\langle \frac{\delta \beta H_f[\zeta,\phi]}{\delta\zeta({\bf r})}\rangle=0.
\end{eqnarray}
In Eq.(\ref{mincond}), $\langle X\rangle$ means  $X$ averaged over the fluctuations $\phi$ with the probability
$\propto \exp(-\beta H_f)$.

As shown in Ref.\cite{ciach:08:1,ciach:11:0}, $\langle \zeta({\bf r})\rangle$ is equal to 
   the ensemble average of  $\hat\zeta({\bf r}')$ for $|{\bf r}'-{\bf r}|<D/2$, further
   averaged over the mesoscopic region around ${\bf r}$ (see Fig.\ref{cartoon}b). 
 Likewise, the correlation function for $\zeta$ at the points ${\bf r}_1$ and ${\bf r}_2$ is equal to 
 the microscopic correlation function  between
 the points ${\bf r}'$ and ${\bf r}''$
  belonging to the mesoscopic regions around  ${\bf r}_1$ and ${\bf r}_2$ respectively, 
  averaged over these regions.    
 
We assume that $U[\zeta]$ is given by the standard expression
\begin{equation}
\label{U}
 U[\zeta]=\frac{1}{2}\int d{\bf r}_1\int d{\bf r}_2 V_{co}({\bf r}_1-{\bf r}_2)\zeta({\bf r}_1)\zeta({\bf r}_2),
\end{equation}
where for the interaction potential $V(r_{12})$ depending only on the distance  $r_{12}=|{\bf r}_1-{\bf r}_2|$,
\begin{eqnarray}
\label{Vco}
 V_{co}(r_{12})= V(r_{12})g(r_{12}).
\end{eqnarray}
We assume that  the microscopic pair distribution function for the volume fraction, $g$,
depends on $r_{12}$, and vanishes for $r_{12}<\sigma$, where $\sigma$ is the particle diameter. In practice,
to determine the structure on the mesoscopic length scale, one may use the approximation $g(r)=0$ or $g(r)=1$ 
for $r<\sigma$ or  $r>\sigma$, respectively.

 We further assume that  the entropy $S$ satisfies the relation
 \begin{equation}
 \label{Fh}
  -TS=F_h=\int d{\bf r} f_h(\zeta({\bf r})),
 \end{equation}
 where 
 $F_h$ is the
 free-energy of the reference hard-sphere system in the local-density approximation.
 The local density approximation is justified in the studies of the
 structure on the mesoscopic length scale. Indeed, the portion of the phase diagram
 obtained in Ref.\cite{edelmann:16:0} in the much more accurate White Bear version of 
 the DFT~\cite{roth:02:0} agrees quite well with the 
 results obtained in the local density approximation in Ref.~\cite{pini:17:0}.
 For the free-energy density of the hard-sphere reference system,  we assume the Percus-Yevick approximation,
\begin{eqnarray}
\label{PY}
\beta f_h(\zeta)=\rho^*\ln(\rho^*)-\rho^*+
\rho^*\Bigg[\frac{3\zeta(2-\zeta)}{2(1-\zeta)^2}-\ln(1-\zeta)\Bigg],
\end{eqnarray}
where $\rho^*=6\zeta/\pi$. Different approximations, such as the Carnahan-Starling approximation, are also possible. 
 
In order to calculate the second term in (\ref{Ome}), we need to make approximations. 
The magnitude of the relevant fluctuations is small ($0\le \zeta({\bf r})+\phi({\bf r})\le \zeta_{cp}$),
and (\ref{Hf1}) can be approximated by a truncated Taylor expansion,
\begin{eqnarray}
\label{Hf2}
 \beta H_f[\zeta,\phi]=\int d{\bf r}\sum_{n=1} \frac{a_n(\zeta({\bf r}))}{n!}\phi({\bf r})^n\\
 \nonumber
 +  \int d{\bf r}_1\int d{\bf r}_2 \Big[\frac{1}{2}\beta V_{co}(r_{12})\phi({\bf r}_1)\phi({\bf r}_2) 
 + \beta V_{co}(r_{12})\zeta({\bf r}_1)\phi({\bf r}_2)\Big]-\beta\mu\int d{\bf r}\phi({\bf r})
\end{eqnarray}
where 
\begin{eqnarray}
\label{fhn}
a_n(\zeta({\bf r}))=\frac{\partial^n\beta f_h (\zeta({\bf r})) }{\partial\zeta({\bf r})^n}.
\end{eqnarray}

We further assume that the periodic order can be destroyed by the fluctuations that vary slowly on the
length scale of the size of the unit cell of the ordered pattern.  
For slowly varying fluctuations we can assume that
within a single unit cell $\phi$ is nearly constant, and we replace $H_f$ (Eq.(\ref{Hf2})) by
\begin{eqnarray}
 \label{Hf3}
\beta \bar H_f[\zeta,\phi]=
 \frac{1}{2}\int \frac{d{\bf k} }{(2\pi)^3}\tilde \phi({\bf k})\beta\tilde V_{co}(k)\tilde \phi(-{\bf k})+\int d{\bf r}\Bigg[
 \sum_{n=2}\frac{A_n[\zeta]}{n!}\phi({\bf r})^n+C_1[\zeta]\phi({\bf r})
 \Bigg],
\end{eqnarray}
where
$C_1[\zeta]=
 A_1[\zeta] +\beta\tilde V_{co}(0)\bar\zeta -\beta\mu$,
we have introduced the functionals
\begin{eqnarray}
\label{Anzeta}
 A_n[\zeta]=\frac{1}{V_u}\int_{V_u} d{\bf r} a_n(\zeta({\bf r})),
\end{eqnarray}
by $\bar\zeta$ we denote the space-averaged volume fraction, 
\begin{eqnarray}
 \bar\zeta=\frac{1}{V_u}\int_{V_u} d{\bf r} \zeta({\bf r}), 
\end{eqnarray}
 $V_u$ is the volume of the unit cell, and  $\tilde f({\bf k})$ 
denotes the function $f$ in the Fourier representation. We use the mixed real-space and Fourier 
representation for convenience.

Let us compare the correlation functions obtained with the effective Hamiltonian (\ref{Hf2}) and (\ref{Hf3}).
In the case of ordered structures, the correlation function 
$\langle \phi({\bf r}_1)\phi({\bf r}_2)\rangle$ obtained with the probability distribution 
$\propto \exp(-\beta H_f)$,
depends on both, ${\bf r}_1$ and ${\bf r}_1-{\bf r}_2$. However, 
when the effective Hamiltonian is approximated by (\ref{Hf3}),
then the correlation function depends only on the distance between the considered points. 
It is instructive to consider 
the Gaussian correlations, with the expansions in (\ref{Hf2}) and (\ref{Hf3}) truncated at $n=2$. 
For simplicity let us assume that the term linear in $\phi$ vanishes. In the Gaussian approximation
the inverse correlation functions are given by the second 
functional derivatives of $\beta H_f[\zeta,\phi]$ and $\beta \bar H_f[\zeta,\phi]$ with respect to $\phi$, 
\begin{eqnarray}
 {\cal C}_0({\bf r}_1,{\bf r}_1-{\bf r}_2):=\frac{\delta^2 \beta H_f}{\delta\phi({\bf r}_1)\delta\phi({\bf r}_2)}=
 \beta V_{co}(r_{12}) + 
a_2(\zeta({\bf r}_1))\delta({\bf r}_1-{\bf r}_2)
\end{eqnarray}
and
\begin{eqnarray}
\label{C0}
C_0({\bf r}_1-{\bf r}_2):=\frac{\delta^2 \beta \bar H_f}{\delta\phi({\bf r}_1)\delta\phi({\bf r}_2)}=
\beta V_{co}(r_{12})+A_2[\zeta] \delta({\bf r}_1-{\bf r}_2)
\end{eqnarray}

 From the above expressions for ${\cal C}_0$ and $ C_0$ and from Eq.(\ref{Anzeta})
 with $n=2$, it follows that $ C_0$ can be obtained from ${\cal C}_0$ by averaging  over 
 the unit cell of the periodic structure.
From now on we consider the approximate theory with the effective Hamiltonian $\bar H_f$ (Eq.(\ref{Hf3})).

 As already noted in the introduction, the standard DFT describes very well the structure of
simple fluids on the microscopic length scale, but fails to predict the correct topology of the
 phase  diagram
in the inhomogeneous, self-assembling systems. We restrict our attention to the latter systems, where
the theory needs to be improved.
The inhomogeneous distribution of particles occurs when $\tilde V_{co}(k)$
takes the global minimum at $k=k_0>0$, $\tilde V_{co}(k_0)<0$ 
and the minimum is deep. For such potentials, we make the approximation
\begin{eqnarray}
\label{Vcoap}
 \beta \tilde V_{co}(k)=\beta^*(-1+v_2(k-k_0)^2)+...\approx \beta^*(-1+\frac{v_2}{4k_0^2}(k^2-k_0^2)^2)+...
\end{eqnarray}
where 
\begin{equation}
\label{v2}
 v_2=\frac{\tilde V_{co}^{''}(k_0)}{2|\tilde V_{co}(k_0)|},
\end{equation}
 and the second equality holds for $k\approx k_0$, i.e. 
for the relevant fluctuations. The density waves with the wavenumber $k_0$ appear with the highest probability,
because $\tilde V_{co}(k)$ takes the minimum for $k=k_0$; fluctuations with $k$ significantly different
from $k_0$ occur with much smaller probability. 
We have introduced the dimensionless temperature
\begin{eqnarray}
\label{T*}
 T^*=1/\beta^*=\frac{k_B T}{|\tilde V_{co}(k_0)|}.
\end{eqnarray}
$T^*$ represents the ratio between the thermal energy, and the energy decrease per unit
volume associated with the excitation
of the volume-fraction wave $\sqrt{2}\cos(k_0z)$ in the homogeneous state (see the first term in (\ref{Hf3})).

 Note that Eq.(\ref{Vcoap})  is not valid in simple systems with purely attractive interactions.
The attractive potential  in Fourier representation 
takes the minimum at $k=k_0=0$, and its expansion about 
the minimum is proportional to $k^2-|const.|$. The wavelength of the most probable 
density wave in simple systems is 
$2\pi/k_0\to \infty$, while in the self-assembling systems $2\pi/k_0$ is finite.
In physical terms, the simple systems tend to a macroscopic separation into the gas and liquid phases, 
because large aggregates of the particles are favoured by the attractive potential. In contrast, 
the systems considered in this work tend to a microseparation into aggregates formed on the mesoscopic length scale,
 because the repulsion suppresses further growth of the clusters. 
In the rest of this work we assume that the interaction potential can be approximated by Eq.(\ref{Vcoap}), 
and only for such
systems the considerations in the rest of this work are valid.

Since for each fixed $\zeta({\bf r})$, the coefficients $A_n[\zeta]$ are just numbers, 
Eq.(\ref{Hf3}) with (\ref{Vcoap}) has
the form similar to the LB functional~\cite{brazovskii:75:0}. When the series in (\ref{Hf3}) 
is truncated at the term 
$\propto \phi^4$, we 
obtain the LB functional.  Thus, we can directly apply the results obtained in the LB theory by the FT methods 
\cite{brazovskii:75:0,ciach:12:0} for 
determination of the explicit form of the second term in (\ref{Ome}). 
In order to calculate this term, we need to know the correlation function 
\begin{equation}
 G({\bf r}_1-{\bf r}_2):=\langle\phi({\bf r}_1)\phi({\bf r}_2)\rangle.
\end{equation}

In the Brazovskii $\varphi^4$ theory, inverse correlation
function $\tilde  C(k)=1/\tilde G(k)$ satisfies the 
self-consistent equation 
(self-consistent Hartree approximation)~\cite{brazovskii:75:0,podneks:96:0,ciach:08:1,ciach:12:0,ciach:11:0}
\begin{eqnarray}
\label{Ck}
 \tilde  C(k)=\tilde  C_0(k)+\frac{A_4[\zeta]}{2}{\cal G}
\end{eqnarray}
where $C_0$ is given in (\ref{C0}), and
\begin{eqnarray}
\label{calG}
 {\cal G}:=\langle \phi({\bf r})\phi({\bf r})\rangle =\int \frac{d{\bf k}}{(2\pi)^3}\tilde G(k).
\end{eqnarray}
 By construction of the mesoscopic theory, the cutoff $2\pi/D$ is present in the integral 
in (\ref{calG}).
When $\tilde V_{co}$ is approximated by (\ref{Vcoap}) and $0\ll k_0\ll2\pi/D$,  then 
the main contribution to $ {\cal G}$ is cutoff-independent, and is given by~\cite{brazovskii:75:0,podneks:96:0,ciach:12:0}
\begin{eqnarray}
\label{Gg}
  {\cal G}\approx 
\frac{2a\sqrt T^*}{Z[\zeta]}
\end{eqnarray}
where 
\begin{eqnarray}
\label{a}
 a=\frac{k_0^2}{4\pi\sqrt v_2},
\end{eqnarray}
\begin{equation}
\label{Z}
 Z[\zeta]:=\sqrt {\tilde C(k_0)},
\end{equation}
and $v_2$ is defined in (\ref{v2}). Note that $a$ characterizes the interaction potential.
We should stress that in this mesoscopic theory, $\langle \phi({\bf r})\phi({\bf r})\rangle$ 
does not represent the microscopic
correlation function calculated at zero distance. It is rather 
the microscopic correlation function between two points
belonging to the same mesoscopic region around ${\bf r}$, and 
averaged over this mesoscopic region~\cite{ciach:08:1,ciach:11:0}. Thus, ${\cal G}$ 
can be considered as a measure of
local deviations from the space-averaged volume fraction $\bar\zeta$.

 Eqs.(\ref{Ck}), (\ref{Gg})  and (\ref{Z}) can be easily solved for $k=k_0$, and $Z[\zeta]$ 
 in the $\varphi^4$ theory is given by the expression~\cite{ciach:12:0}
\begin{eqnarray}
\label{sqrtr}
 Z[\zeta]= \frac{W[\zeta]}{6}+\frac{2(A_2[\zeta]-\beta^*)}{W[\zeta]},
\end{eqnarray}
 where 
 \begin{eqnarray}
 \label{W}
  W[\zeta]=\Bigg[
108 A_4[\zeta] a\sqrt {T^*} +12\sqrt{(9 aA_4[\zeta])^2T^*-12 (A_2[\zeta]-\beta^*)^3}
\Bigg]^{1/3}.
 \end{eqnarray}

In order to evaluate the fluctuation contribution to $\Omega[\zeta]$,
 we decompose $H_{f}[\zeta,\phi]$ into two parts~\cite{landau:58:0,ciach:08:1,ciach:12:0}
\begin{equation}
\label{Har}
H_{f}[\zeta,\phi] ={\cal
H}_{G}[\zeta,\phi]+\Delta{\cal H}[\zeta,\phi],
\end{equation}
where
\begin{eqnarray}
\label{Haga}
 {\cal H}_G[\zeta,\phi]=
\frac{1}{2}\int \frac{d{\bf k}}{(2\pi)^3}\tilde \phi({\bf k})\tilde C(k)\tilde\phi(-{\bf k}).
\end{eqnarray}
Assuming $\Delta{\cal H}\ll {\cal H}_{G}$,
we obtain \cite{landau:58:0,ciach:08:1,ciach:12:0}
\begin{eqnarray}
\label{OmHarval} 
\beta\Omega[\zeta]\approx
\beta\Omega_{co}[\zeta]-\log\int D\phi \, e^{-\beta{\cal H}_G}
+\langle \beta\Delta{\cal H}\rangle_G +O(\langle \beta\Delta{\cal
H}\rangle_G^2),
\end{eqnarray}
where $\langle ...\rangle_G$ denotes averaging with the Gaussian Boltzmann
factor $\propto e^{-\beta{\cal H}_G}$.
The fluctuation contribution
in Eq.~(\ref{OmHarval}) for the approximations (\ref{Ck}), (\ref{sqrtr}) was calculated in
Ref.~\cite{brazovskii:75:0,podneks:96:0,ciach:08:1,ciach:12:0}, 
and the final expression for $\beta\Omega[\zeta]$ in the $\varphi^4$
theory is 
\begin{equation}
 \label{OmHarval1}
\beta\Omega[\zeta]/{\cal V}\approx \beta\Omega_{co}[\zeta]/{\cal V}+2a\sqrt{T^*}Z[\zeta]-
\frac{A_4[\zeta]a^2 T^*}{2Z[\zeta]^2},
\end{equation}
where by ${\cal V}$ we denote the volume of the system,  $\beta\Omega_{co}[\zeta]$ is given 
in (\ref{Omco})-(\ref{Fh}), and
$T^*$, $a$,  $A_n[\zeta]$ and $Z[\zeta]$ are given in (\ref{T*}), (\ref{a}), (\ref{Anzeta}), 
and (\ref{sqrtr})-(\ref{W})
The expression (\ref{OmHarval1}) for the grand potential can be minimized numerically 
to yield the equilibrium structure in the 
presence of mesoscopic fluctuations for any value of the chemical potential $\mu$ and temperature $T$.

In Eq.(\ref{OmHarval1}), the fluctuation contribution has been obtained under
many assumptions and approximations. In particular,
the expansion in (\ref{Hf3}) has been truncated at $n=4$. 
This is justified when the higher-order terms are negligible for 
the dominant fluctuations. 
When the expansion in (\ref{Hf3}) is truncated at $n=6$,
then on the same level of the self-consistent one-loop  approximation in the $\varphi^6$ theory 
we obtain~\cite{ciach:06:2,patsahan:07:0}
\begin{eqnarray}
\label{Ck6}
 \tilde  C(k_0)=A_2[\zeta]-\beta^*+\frac{A_4[\zeta]}{2}{\cal G}+\frac{A_6[\zeta]}{8}{\cal G}^2.
\end{eqnarray}
Using (\ref{Ck6}),  (\ref{Gg}) and (\ref{Z}),
we obtain the equation for $Z[\zeta]$ in the $\varphi^6$-theory,
\begin{eqnarray}
\label{Y}
 Z[\zeta]^4-(A_2[\zeta]-\beta^*)Z[\zeta]^2-
 a\sqrt T^*A_4[\zeta]Z[\zeta]
 -\frac{a^2 T^*A_6[\zeta]}{2}=0
\end{eqnarray}
and the expression for the grand potential (see (\ref{Har})-(\ref{OmHarval})),
\begin{equation}
 \label{OmHarval2}
\beta\Omega[\zeta]/{\cal V}\approx \beta\Omega_{co}[\zeta]/{\cal V}+2a\sqrt{T^*}Z[\zeta]-
\frac{A_4[\zeta]a^2 T^*}{2Z[\zeta]^2}-\frac{A_6[\zeta]a^3 T^{*3/2}}{3Z[\zeta]^3}.
\end{equation}
Eqs.(\ref{OmHarval2}) and (\ref{Y}) with (\ref{Omco}), (\ref{U})-(\ref{PY}), (\ref{T*}), 
(\ref{a}) and (\ref{Anzeta}) are
the main result of this section. Minimization of $\beta\Omega[\zeta]/{\cal V}$ gives $\zeta({\bf r})$ 
corresponding to a stable or a metastable phase.

\section{the case of weak order}
\label{weak}
The average volume fraction in the ordered phase can be written in the form 
\begin{equation}
 \zeta({\bf r})=\bar\zeta +\Phi({\bf r}),
\end{equation}
where by definition of $\bar \zeta$, $\Phi$ must satisfy $\int d{\bf r}\Phi({\bf r})=0$. 
In the ordered periodic phases 
\begin{eqnarray}
\label{shells}
\Phi({\bf r})=\sum_{n\ge 1}\Phi_n g_n({\bf r}),
\end{eqnarray}
where 
$g_n({\bf r})$ represent orthonormal basis functions for the $n$-th 
shell that have the symmetry of the considered phase, 
and satisfy the normalization condition~\cite{podneks:96:0}
\begin{eqnarray}
\label{gnorm}
\frac{1}{V_u}\int_{V_u} d{\bf r} g_n({\bf r})^{2}=1.
\end{eqnarray}
By weak order we mean the structure with 
$\Phi({\bf r})$ that can be approximated by the first shell in (\ref{shells}), and has a small magnitude. 
The functions $g_1({\bf r})$ are given by a superposition of plane waves with the wavevectors ${\bf k}_0^j$ 
such that $|{\bf k}_0^j|=k_0$. In Fourier representation
\begin{equation}
\label{g}
\tilde g_1({\bf k})=\frac{ (2\pi)^{d}}{\sqrt{2n}}
\sum_{j=1}^{n}\Big(w\delta({\bf k}-{\bf k}^j_{0})+w^*
\delta({\bf k}+{\bf k}^j_{0})\Big),
\end{equation}
where $ww^*=1$ and $2n$ is the number of the vectors ${\bf k}^j_{0}$ in the first
shell.
In the case of the lamellar phase with the oscillations in direction $\hat{\bf z}$, $g_1(z)=\sqrt 2 \cos(k_0z)$.
The expressions for $g_1({\bf r})$ for the remaining phases can be found in Appendix.
In the one-shell approximation we denote the amplitude by $\Phi$ (we omit the subscript 1). 
By definition of $\bar\zeta$, $\int_{V_u}d{\bf r}g_1({\bf r})=0$. 

 We can  Taylor expand $a_n (\zeta({\bf r}))$ defined in (\ref{fhn}), 
\begin{eqnarray}
\label{fn+m}
 a_n(\zeta({\bf r}))=a_n(\bar \zeta)+\sum_{m\ge 1}\frac{a_{n+m}(\bar \zeta)}{m!}\Phi({\bf r})^m,
\end{eqnarray}
 and the expansion can be truncated in the case of  $\Phi\ll 1$.
Eq.(\ref{fn+m}) with $n=0$,  and Eq.(\ref{g}) lead to an approximate form of $\Omega_{co}$
that at this level of approximation becomes a function of $\bar\zeta$ and $\Phi$ of the form
\begin{eqnarray}
\label{MF}
\beta\Omega_{co}(\bar\zeta,\Phi)/{\cal V}= \beta\Omega_{co}(\bar\zeta)/{\cal V}-\frac{\beta^*}{2}\Phi^2
+\sum_{n\ge 2}^M\frac{a_n(\bar\zeta)}{n!}\kappa_n\Phi^n
\end{eqnarray}
where  $M\ge 4$, 
\begin{equation}
 \beta\Omega_{co}(\bar\zeta)/{\cal V}=\frac{1}{2}\beta^* v_0\bar\zeta^2+\beta f_h(\bar\zeta)-
\beta\mu\bar\zeta,
\end{equation}
\begin{equation}
\label{v0}
 v_0= \tilde V_{co}(0)/|\tilde V_{co}(k_0)|,
\end{equation}
and $\beta^*$ and $a_n(\zeta)$ are defined in (\ref{T*}) and (\ref{fhn}), respectively. 
We have introduced the geometric factors 
\begin{eqnarray}
\label{kappa}
 \kappa_n=\frac{1}{V_u}\int_{V_u} d{\bf r}g_1^{n}({\bf r})
\end{eqnarray}
that except from $\kappa_2=1$ take different values for different phases, and are given in Appendix.
Here we limit ourselves to the $\varphi^M$ theory with $M=4$ and $M=6$. 

For $n\ge 1$ the expansion (\ref{fn+m}) leads to approximate forms of $A_n[\zeta]$ defined in (\ref{Anzeta}) 
that also become functions
of $\bar\zeta$ and $\Phi$, and inserted in (\ref{Hf3}) lead to $\bar H_f$ that is a function of 
$\bar\zeta$ and $\Phi$, and 
a functional of the fluctuation $\phi$. 
In the case of weak order, where fluctuations play an important role,
we may expect that $\Phi$ and the dominant fluctuations 
are of the same order of magnitude. Thus, in the $\varphi^M$-theory we keep in $\bar H_f$ only terms
$\Phi^n\phi^m$ with $n+m\le M$. In this case, 
$A_n[\zeta]$ in Eq.(\ref{Hf3})
is approximated by 
\begin{eqnarray}
\label{An}
 A_n(\bar\zeta,\Phi)=a_n(\bar\zeta)+
 \sum_{m=2}^{M-n} \frac{a_{n+m}(\bar\zeta)\kappa_{m}\Phi^{m}}{m!},
\end{eqnarray}
 and the  series in Eq.(\ref{Hf3}) is truncated at $n=M$.

An important consequence of the reduction of the functionals $\Omega_{co}[\zeta]$, $A_n[\zeta]$ 
to the functions of the two variables,
$\bar \zeta$ and $\Phi$, is the reduction of the equilibrium condition (\ref{mincond}) to just two equations,
\begin{eqnarray}
\label{eqP}
 \frac{\partial \beta\Omega_{co}(\bar\zeta,\Phi)}{\partial \Phi}+
 \langle \frac{\partial \beta \bar H_f}{\partial \Phi}\rangle=0
\end{eqnarray}
and
\begin{eqnarray}
\label{eqz}
 \frac{\partial \beta\Omega_{co}(\bar\zeta,\Phi)}{\partial \bar\zeta}
 +\langle \frac{\partial \beta \bar H_f}{\partial \bar\zeta}\rangle=0.
\end{eqnarray}
The explicit forms of Eqs.(\ref{eqP}) and (\ref{eqz}), and the explicit
expression for the thermodynamic pressure $p=-\Omega/V$ (Eq.(\ref{OmHarval1}) or (\ref{OmHarval2})) are given in Appendix.
From these equations, we can obtain $\Phi$, $\mu^*=\mu/|\tilde V_{co}(k_0)|$  and $p^*=p/|\tilde V_{co}(k_0)|$ 
for the stable or metastable phase 
 for each value of $\bar\zeta$, and for each set of the geometric factors $\kappa_n$
characterizing the considered phases. From   $\mu^*(\bar\zeta)$ and  $p^*(\bar\zeta)$,
we obtain $p^*(\mu^*)$ isotherms by eliminating $\bar \zeta$.
The phase coexistence occurs when the lines $p^*(\mu^*)$ for two phases intersect. 
In addition, we obtain the EOS isotherms $p^*(\bar\zeta)$.

Solving the algebraic equations (see Appendix) is an easier task than finding the minimum of the functional in
Eq.(\ref{OmHarval1}) or (\ref{OmHarval2}). Unfortunately, because of the one-shell approximation,
these equations
are valid only for nearly sinusoidal shapes of the volume-fraction profiles.
Such shapes were found in the MF-type
DFT theory~\cite{pini:17:0} only
at the high-$T$ part of the phase diagram.
The volume-fraction profiles deviate significantly from sinusoidal shapes in 
a substantial part of the phase diagram~\cite{pini:17:0}, therefore the
one-shell
 approximation developed in this section is certainly an oversimplification,
 except at relatively high $T$, where the order is weak
 and the fluctuations are strong. 
 
 \section{Examples}
 \label{examples}
 In this section we consider hard spheres that for distances larger than $\sigma$ interact with the 
  SALR potential of the form of a square well followed by a repulsive ramp,
 since for this potential the phase diagram has been obtained in MC simulations~\cite{zhuang:16:0,zhuang:16:1}. 
 In Ref.\cite{zhuang:16:0,zhuang:16:1}, the potential between the particles has the form
 \begin{equation}
\label{u}
u(r) = \left\{ \begin{array}{ll}
\infty& \textrm{if $r < 1$}, \;\; \;\\
-\epsilon & \textrm{if $1<r < 3/2$}, \;\; \;\\
\epsilon\xi(\kappa -r) & \textrm{if $3/2<r<\kappa$}, \;\; \;\\
0 & \textrm{if  $r > \kappa$,}
\end{array} \right.
\end{equation}
 where  the  length unit   is the particle diameter $\sigma$. The depth of the 
square well, $\epsilon$,  sets the unit of energy ($\epsilon=1$ is assumed),
and the dimensionless temperature is defined by
$\bar T=k_BT/\epsilon$. 

We shall focus on two systems, System 1 with $\xi=0.05$ and $\kappa=4$,
 and System 2, with  $\xi=6$ and $\kappa=2$. In System 1, the slope of the ramp is small, and
 the weak repulsion has a relatively large range, while in System 2, the strong repulsion is of a short range. 
 In Ref.\cite{zhuang:16:1}, no stable periodic structures were found for System 2. In System 1, 
 the simulations were restricted
 to $\rho^*<0.45$ ($\zeta<0.236$). For this range of dimensionless densities, 
 the D, C, H and L phases (see Fig.\ref{Phases}) are
 stable for increasing density 
 for $\bar T<\bar T_C\approx 0.41$. There is also 
 a narrow window of stability of the double-gyroid phase 
 between the H and L phases at low $\bar T$. For  $0.41\approx \bar T_C<\bar T< \bar T_H\approx 0.49$, the sequence of phases is D,  H and L.
 The H phase disappears for $\bar T>\bar T_H\approx 0.49$. For $0.49\approx\bar T_H<\bar T< \bar T_L\approx 0.535$ 
 and $\rho^*< 0.45$,
 L is the only stable ordered phase. For a small range of temperature below 
 $\bar T_L\approx 0.535$, a reentrant melting of the L phase was observed, i.e. 
 as long as  $\rho^*< 0.43$, the sequence
 of phases for increasing density is D, L, D.

\begin{figure}[h]
 \centering
 \includegraphics[scale=0.32]{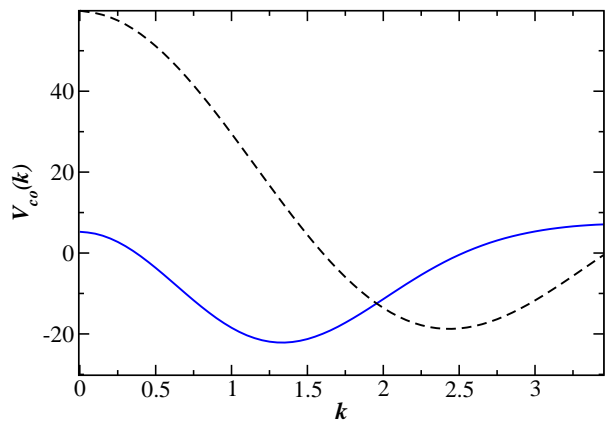}
\caption{ $V_{co}$ defined in Eq.(\ref{Vco}) in Fourier representation. The solid line corresponds to System 1
 (the interaction potential (\ref{u})
with $\kappa=4$ and $\xi=0.05$),
and the dashed line corresponds to System 2  (the interaction potential (\ref{u})
with $\kappa=2$ and $\xi=6$). The wavenumber $k$ is in $\sigma^{-1}$ units, and 
$\tilde V_{co}(k)$ is in $\epsilon$-units, with $\sigma$ and $\epsilon$ denoting the particle diameter and
the depth of the square-well (Eq.(\ref{u})). }
\label{VcoFig}
\end{figure}

Note that in the expression for the internal energy (Eq.(\ref{U})) we use volume 
fractions instead of densities, therefore the potential $u$ should be rescaled,
i.e. in Eq.(\ref{U}), $V(r)=(6/\pi)^2u(r)$. Moreover, we consider 
 the product $V_{co}(r)=V(r)g(r)$,
where $g(r)=0$ for $r<1$ and  $g(r)=1$ for $r>1$ (in $\sigma$-units). The function $V_{co}$ 
for the two considered systems is shown in Fig.\ref{VcoFig}
in Fourier representation.

The period of the most probable density
wave is $2\pi/k_0$, where $\tilde V_{co}(k)$ takes the minimum at $k_0$. 
Fig.\ref{VcoFig} shows that the excitation of the most probable density wave 
leads to a similar energy gain per unit volume in the two systems. 
From the energy point of view, the two systems should show
similar tendency for periodic ordering. Simulations, however, show periodic ordering only in System 1. 

Apart from  $\tilde V_{co}(k_0)$ and $k_0$, the relevant parameters characterizing the potential in 
our theory are $v_0$ and $a$
defined in Eq.(\ref{v0}) and in Eq.(\ref{a}), respectively. For the considered potentials we have: 
\begin{equation}
{\rm  System \hskip 0.2cm 1:} \hskip1cm \kappa=4, \hskip0.9cm \xi=0.05, \hskip0.9cm k_0\approx 1.3,\hskip0.9cm v_0\approx 0.2345,\hskip0.9cm a\approx 0.116,
\end{equation}
\begin{equation}
{\rm   System \hskip 0.2cm 2:} \hskip1cm \kappa=2, \hskip1cm \xi=6, \hskip1cmk_0\approx 2.435, \hskip1cm v_0\approx 3.18, \hskip1cm a\approx 0.566 ,
\end{equation}

The larger value of $k_0$ in System 2 leads to a value of $a$ almost 5 times larger than in System 1. 
Note that the fluctuation contributions 
to the grand potential are proportional to $(a^2 T^*)^{n/2}$ with $n=1-3$ (see (\ref{OmHarval2})). Thus, at given 
$T^*$ the fluctuation contribution in System 2 is expected to be larger than in System 1.
However, the increase of $a$
can be compensated by a decrease of $T^*$ to obtain in System 2 the same value of the parameter $(a^2 T^*)^{n/2}$
as in System 1. This simple analysis indicates that the ordered phases should occur in System 2, but at much lower
temperature than in System 1. Physically, the larger period means a smaller number of aggregates per unit volume, 
and smaller entropy associated with distribution of these aggregates in space.
For this reason the disordering effect of entropy in System 1
is weaker than in System 2, and in the former the ordered phases can be stable at higher 
temperature than in the latter.

\begin{figure}[h]
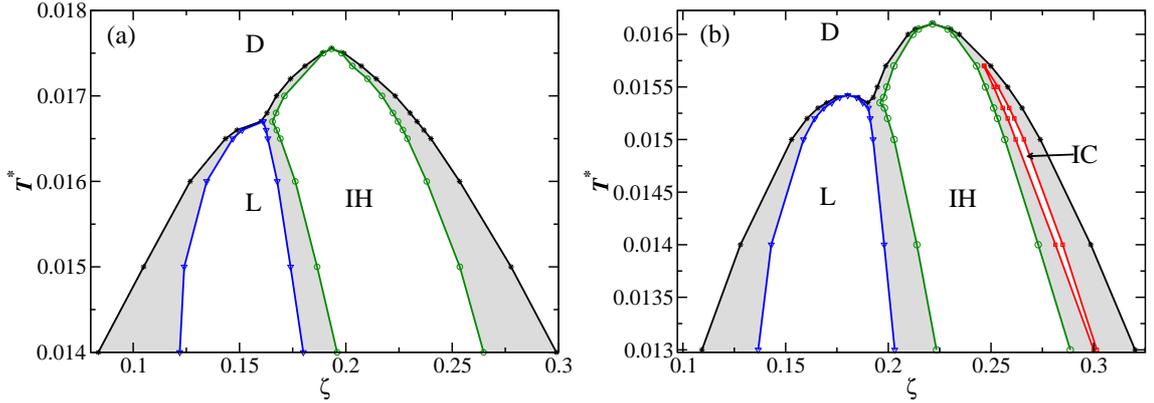

 \centering
 \includegraphics[scale=0.3]{fig4a.eps}
  \includegraphics[scale=0.3]{fig4b.eps} 
\caption{The high-$T^*$ part of the phase diagram in System 1  (the interaction potential (\ref{u})
with $\kappa=4$ and $\xi=0.05$)
 in the $\varphi^4$-theory (a)
and in the  $\varphi^6$-theory (b).
The reduced temperature 
$T^*$ is defined in Eq.(\ref{T*}), and the volume fraction $\zeta$ is dimensionless.
To compare with the phase diagram obtained in simulations, note that  dimensionless temperature and density 
in Ref.\cite{zhuang:16:0}
are $\bar T=T^*|\tilde V_{co}(k_0)| \approx 22.14 T^*$
and  $\rho^*=6 \bar \zeta/\pi$.
D, L, IH, IC denote the disordered, lamellar, inverted hexagonal and inverted cubic (bcc) phases (Fig.\ref{Phases}).
The  two-phase coexistence regions are gray-shaded. The symbols indicate the values of $T^*$
for which the phase coexistence was calculated according to Eqs.(\ref{P})-(\ref{pres}).}
\label{PhDiag}
\end{figure}

In order to obtain the phase diagrams in the two systems, we perform the analysis described in sec.~\ref{weak}. 
For the reference-system free-energy density we assume the PY approximation (\ref{PY}). 
The derivatives  $a_n(\bar\zeta)$ of $\beta f_h(\bar\zeta)$ can be easily calculated.
We solve (\ref{P})-(\ref{pres}) (with
$Z(\bar\zeta,\Phi)$  given by (\ref{sqrtr}) and (\ref{W}) in the $\varphi^4$-, or by (\ref{Y}) 
in the $\varphi^6$-theory, with $A_n[\zeta]$ 
approximated by  (\ref{An})). In some cases
 there is more than one solution for $\Phi$. We have verified that the larger value of 
 $\Phi$ leads to larger $p^*$ 
 for given $\mu^*$; we have selected this solution, and obtained $\mu^*(\bar\zeta)$ and $p^*(\bar\zeta)$
for each set of the geometric factors $\kappa_n$ and for fixed $T^*$. Finally, 
from the intersections of the isotherms $p^*(\mu^*)$,
we have obtained the phase diagrams.
The results for the high-temperature part of the 
phase diagram in System 1 in the $\varphi^4$- and  $\varphi^6$-theory are shown in Fig.\ref{PhDiag}.
\vskip1cm
The main features of the phase diagram in the  $\varphi^4$-
and  $\varphi^6$-theory are similar, but the details are different. In the $\varphi^6$-theory 
the IC phase is stable,
whereas in the  $\varphi^4$-theory it is only metastable for the volume fraction inside
 the IH-D two-phase region. 
However, the difference between the grand potentials in the stable IH, D and metastable IC phases is very small,
therefore the accuracy of the approximation plays a significant role in determining the stability of the IC phase. 
 In the  $\varphi^6$-theory the ordered phases are stable for lower temperature and larger volume fraction
 than in the  $\varphi^4$-theory.
The L phase is stable for the volume fractions that in the  $\varphi^6$-theory agree pretty well with simulations.  
Another difference between the two approximations is the reentrant melting of the L phase at high temperature,
present only in the  $\varphi^6$-theory.
  Recall that in simulations,  the sequence of phases D,L,D was found close to $\bar T=\bar T_L$,
therefore we conclude that the shape of the high-T part of the phase diagram is correctly reproduced by our 
$\varphi^6$-theory, at least for $\zeta<0.236$.

Our temperature scale is different than in simulations (see Eq.(\ref{T*})), and the relation is 
$T^*=\bar T/|\tilde V_{co}(k_0)|$, i.e. $T^*\approx \bar T/22.14$ in System 1. 
While the range of volume fraction corresponding to the stability of the L phase in theory and simulations is in 
rather good agreement, the temperature range of stability of the L phase in our theory is smaller than
in simulations;
the L phase looses stability in System 1 at $T^*_L\approx 0.0154$, that corresponds to $\bar T_L\approx 0.34$,
whereas simulations give $\bar T_L\approx 0.535$. On the one hand, 
the fluctuation contribution in our approximation may be overestimated.
On the other hand, in simulations the finite size of the system and periodic boundary conditions 
 suppress the mesoscopic fluctuations 
of large wavelengths 
that destroy the periodic order in the bulk. For this reason,  in simulations the temperature 
range corresponding to 
stability of the periodic structures may be overestimated. 

\begin{figure}[h]
 \centering
 \includegraphics[scale=0.31]{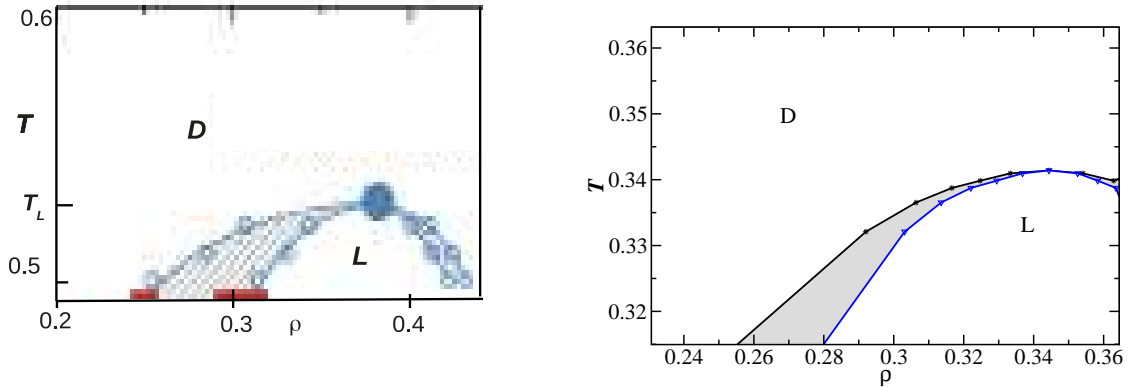}
  \includegraphics[scale=0.3]{fig5b.eps} 
\caption{ The high-T part of the phase diagram in dimensionless density $\rho$ and  temperature 
in $\epsilon$ units for System 1. Panel (a):  the part of the phase diagram obtained in simulations,
Ref.\cite{zhuang:16:0}.
Panel (b): this theory. Since the density range where the IH and IC phases are stable
was not studied in simulations, we do not show the part of the phase diagram corresponding
to stability of these phases. 
Note that in the $\varphi^6$ theory the stability region of the ordered phases is shifted to higher
densities compared
to the $\varphi^4$ theory (see Fig.\ref{PhDiag}). We may expect that in the $\varphi^8$ or higher order theory this trend will
lead to a still better agreement with simulations. Note the coexistence of the L phase with the D 
phase and the reentrant melting
close to $T_L$ 
in both cases, 
in contrast to the coexistence of the L phase with the H and IH phases up to $T_L$ in MF (Fig.\ref{Phases}). 
The different temperature range of the stability of the ordered phases is discussed in the main text. }
\label{part}
\end{figure}

 Note that in MF (Fig.\ref{Phases}),
all the ordered phases are stable up to
 $\bar T_L\approx 2$  (to get
 $k_BT$ in the
 $\epsilon$ units,
$T^*$ in Fig.\ref{Phases} should be multiplied by 22.14), and at
this maximum temperature the density region of all the ordered phases shrinks to $\rho\approx 0.25$. 
This is in a sharp contrast to both, 
our theory and  simulation results,
where at high temperature  the H and C phases are not stable, 
and the low-density D phase coexists with the L rather than with the C phase.
For a better comparison between our theory and simulations, we show in Fig.\ref{part} the high-T part 
of the phase diagram obtained in simulations~\cite{zhuang:16:0} and in this theory.

The large stability region of the IH phase is rather surprising, but since the simulations in Ref.\cite{zhuang:16:1}
were restricted to  $\zeta<0.236$, we cannot verify if our predictions are correct for large volume fractions.
We can only note that the inverse phases,
with periodically distributed voids, have been investigated in simulations in Ref.\cite{lindquist:16:0,lindquist:17:0}. 
Unfortunately,  in Ref.\cite{lindquist:16:0,lindquist:17:0} the phase diagram was not determined.

For the temperature range shown in Fig.\ref{PhDiag}, the phases C and H are not stable. 
The C phase is not even metastable for $T^*>0.011$, and the H-phase is not metastable for $T^*>0.0133$. 
Unfortunately, for $T^*<0.013$ the  one-shell approximation leads to unphysical results for the lamellar phase. 
We obtain the amplitude leading to local volume fractions $\zeta({\bf r})$ that in some regions are negative, 
and in some other regions much greater than one. As shown in Ref.\cite{pini:17:0}, 
$\zeta({\bf r})$ deviates strongly 
from the sinusoidal shape at low $T^*$. Our results indicate that for   $T^*<0.013$
the approximation developed in sec.\ref{weak} is a significant oversimplification, 
and for such temperatures one has to go beyond the one-shell approximation.  
The H phase becomes metastable for  $T^*<0.864T^*_L$, and the C phase becomes metastable for $T^*<0.714 T^*_L$. 
In simulations, the H and C phases become stable for $\bar T<0.916\bar T_L$ and $\bar T<0.766\bar T_L$ respectively.
These temperature ratios in the theory and in simulations are similar. 
Since by decreasing $T^*$ we obtain
the metastable H phase and next the metastable C phase 
(both more stable than the D phase for some temperature interval), we may expect that with  
the proper  shape of the volume-fraction profile of the L phase, i.e. beyond the one-shell approximation,
 the correct low-$T$ part of the  phase diagram can be obtained by a numerical minimization of 
 the functional (\ref{OmHarval2}).

\begin{figure}[h]
 \centering
 \includegraphics[scale=0.3]{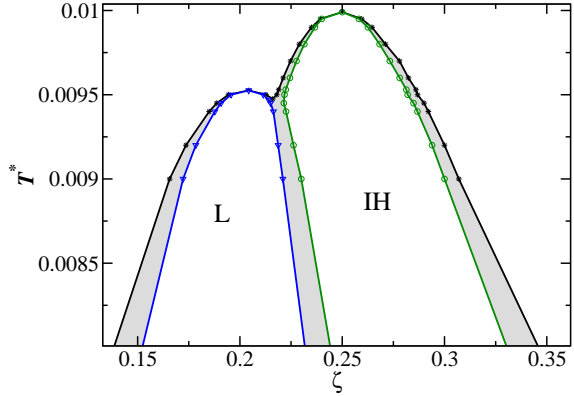}
\caption{The high-$T^*$ part of the phase diagram in System 2 
(the interaction potential (\ref{u}) with $\kappa=2$ and $\xi=6$)
 in   $\varphi^6$-theory. The reduced temperature 
$T^*$ is defined in Eq.(\ref{T*}), and the volume fraction $\zeta$ is dimensionless. 
Note that temperature
in Ref.\cite{zhuang:16:0}
is $\bar T\approx 18.76 T^*$.}
\label{PhDiag2}
\end{figure}

The phase diagram in System 2 has been obtained in the  $\varphi^6$-theory, 
and is shown in Fig.\ref{PhDiag2}. 
The shape of the phase diagram in both systems is similar, except that the IC phase in System 2 is only metastable.
Note, however, that the ordered phases in System 2 are stable at much lower temperatures than in System 1,
in agreement with the simple arguments discussed above. The relation between the temperature scales 
in our theory and in simulations  in System 2 is $T^*=\bar T/|\tilde V_{co}(k_0)\approx \bar T/18.76 $ 
(see Fig.\ref{VcoFig}).

The ratio between the temperature $ T^*_L$ in System 2 and in System 1  in our theory is $0.52$. 
Assuming that in simulations of Ref.\cite{zhuang:16:0} this 
ratio is similar, we estimate the boundary of stability of the L phase in simulations of
System 2 for $\bar T_L\sim 0.27$. In Ref. \cite{zhuang:16:0}, the simulations were performed for $\bar T>0.25$
and $\rho^*<0.55$~\cite{charbo:18:0}, 
therefore if the ordered phases are present in System 2 for  $\bar T<0.25$, they could  not be
detected in these simulations. Thus, there is no contradiction between our predictions and simulations in
Ref.~\cite{zhuang:16:0}.

\begin{figure}[h]
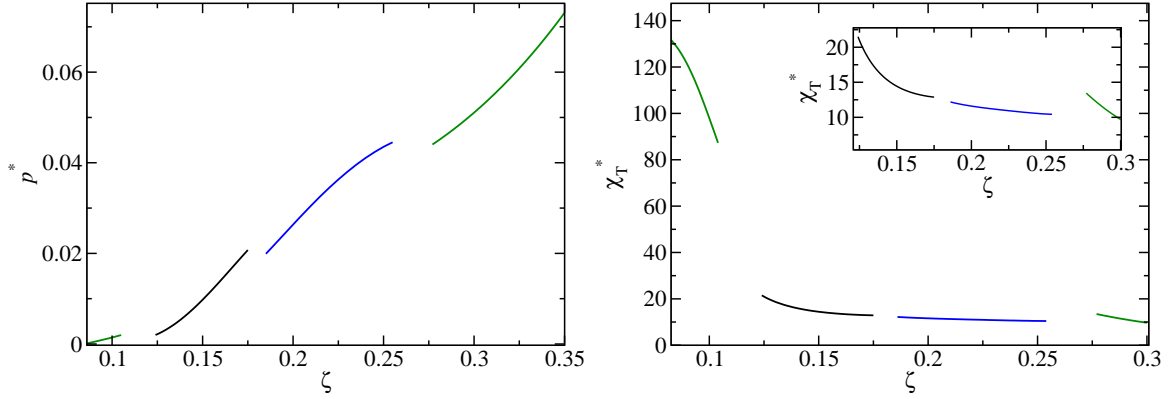

 \centering
 \includegraphics[scale=0.3]{fig7a.eps}
  \includegraphics[scale=0.3]{fig7b.eps}
\caption{The $T^*=0.015$ isotherm for the pressure (a) and compressibility 
$\chi_T^*$ (b) in the $\varphi^4$-theory,
as  functions
of the particle volume fraction for System 1 (model (\ref{u}) with $\kappa=4$ and $\xi=0.05$).
The segments from left to right correspond to the D, L, IH and again
D phases, and are separated by two-phase regions. $p^*$ and $1/\chi_T^*$ are in 
$|\tilde V_{co}(k_0)|/\sigma^{3}$ units.
}
\label{EOS15}
\end{figure}

\begin{figure}[h]
 \centering
 \includegraphics[scale=0.3]{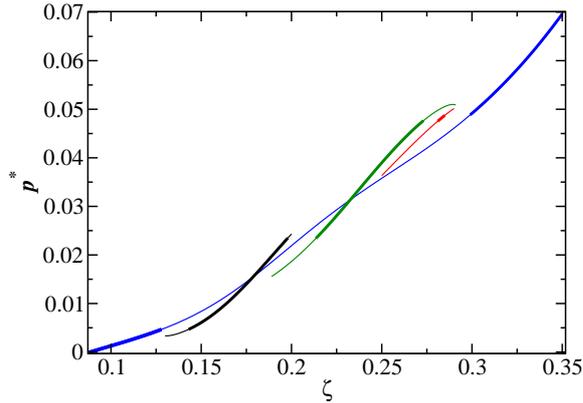}
\caption{The $T^*=0.014$ isotherm for the pressure  as a function
of the particle volume fraction in the $\varphi^6$-theory for System 1 
(model (\ref{u}) with $\kappa=4$ and $\xi=0.05$).
The thick segments from left to right correspond to the stable  D, L, IH, IC  and again
the D phases, and the thin continuations of the thick lines represent the corresponding metastable phase.}
\label{EOS14}
\end{figure}

Another interesting question concerning the SALR systems is  the
effect of self-assembly and periodic ordering of clusters or voids
on the EOS and mechanical properties such as the compressibility
$\chi_T^*=\bar\zeta^{-2}\partial \bar\zeta/\partial \mu^*$. 
This question has been much less studied than the phase diagram
~\cite{ciach:12:0,ciach:16:0,ciach:16:1}.
 We have calculated $p^*(\bar\zeta)$ and $\chi_T^*(\bar\zeta)$ for weakly ordered systems
in the framework of the theory developed in sec.\ref{weak}.  
The $T^*=0.015$ and $T^*=0.014$ isotherms 
for System 1 are shown in Fig.\ref{EOS15} and in Fig.\ref{EOS14} in the $\varphi^4$- and $\varphi^6$-theory,
respectively. Note the characteristic shape of the $p^*(\bar\zeta)$ lines that consist of 
segments with a large slope
separated by the narrow two-phase regions. 
In the periodic
phases the slopes of $p^*(\bar\zeta)$ are larger  than in the metastable D phase 
for the same volume-fraction interval (Fig.\ref{EOS14}). 
As a result, the compressibility in the ordered phases is very low, despite relatively low density. 
In particular, at the D-L phase-coexistence
the compressibility of the L phase is about 4 times smaller than the compressibility of the D phase,
even though the volume fraction in the L phase is not much larger than in the D phase. Even more surprising is 
the larger compressibility in the D phase than in the coexisting IH phase, despite larger density in the former.
Our results show that it is the periodic structure that makes the system  quite stiff, despite relatively large
 volume available for the particles.

\begin{figure}[h]
 \centering
 \includegraphics[scale=0.3]{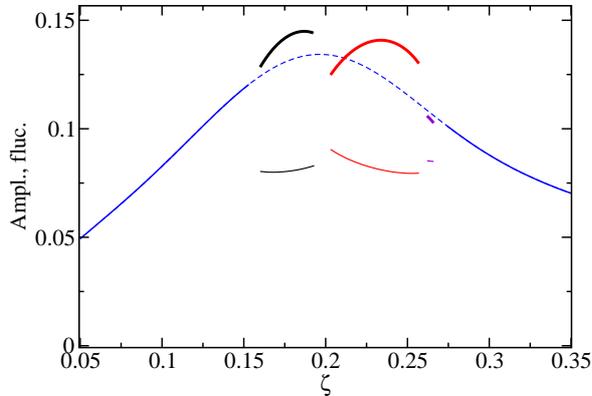}
\caption{The short thick solid lines represent the amplitude $|\Phi|$ 
of the density oscillations for the L, IH and IC
phases 
(from the left to the right)  in the $\varphi^6$-theory for $T^*=0.015$. The short
thin lines represent the mesoscopic fluctuation
$\sqrt{\langle \phi({\bf r})^2\rangle}$ in the corresponding ordered phase. 
The long blue line represents the fluctuation  $\sqrt{\langle \phi({\bf r})^2\rangle}$ in the disordered phase.
The solid and dashed parts refer to the volume-fraction range where the  D phase is stable and
metastable, respectively. }
\label{PHI2}
\end{figure}

In derivation of the approximate form of $\bar H_{f}$ in sec.\ref{weak} (see Eqs. (\ref{Hf2}), (\ref{An})), 
we have assumed that  in the case of weak order the amplitude $\Phi$ of the oscillation of the average 
volume fraction, and the dominant fluctuation
$\phi$ are of the same order of magnitude. To verify this assumption, we  plot $|\Phi|$ and 
$\sqrt{\langle \phi({\bf r})^2\rangle}$ in Fig.\ref{PHI2} for $T^*=0.015$ and a range of $\bar\zeta$.
We can see that in the L, IH and IC phases, $\sqrt{\langle \phi({\bf r})^2\rangle}$ is smaller than $\Phi$ by
a factor  $\approx 1/2$. Thus, for the high-$T^*$ part of the phase diagram, 
this assumption is valid. In addition, in Fig.\ref{PHI2} we plot $\sqrt{\langle \phi({\bf r})^2\rangle}$
in the stable and metastable D phase. Interestingly, $|\Phi|$ in the ordered phase is very similar to 
$\sqrt{\langle \phi({\bf r})^2\rangle}$ in the metastable D phase for the same volume fractions. 
Note that in the D phase, $\sqrt{\langle \phi({\bf r})^2\rangle}$ can be interpreted as the standard
deviation of the local volume fraction in a mesoscopic region from the
space-averaged value $\bar\zeta$. It is a measure of the excess number of particles
in the dense regions, or of depletion of the particles in the  dilute regions.
Our results show that the local structure
in the metastable D and in the stable ordered phases is very similar. Rather large fluctuations in the 
ordered phases mean a large number of defects in the periodic structure. In experiment, it may be difficult
to distinguish the ordered and the disordered phases in the case of weak order.

\section{discussion and summary}
\label{discussion}
The main result of this work is the ``user-friendly'' expression for the grand-potential functional
of the volume fraction of particles, Eqs.(\ref{OmHarval1}) with (\ref{sqrtr}),
or (\ref{OmHarval2}) with (\ref{Y}) that
can be directly minimized numerically. In contrast to standard DFT functionals, our formula contains a
contribution from mesoscopic fluctuations. This contribution has been obtained within the well-known 
field-theoretic formalism on the level of the Brazovskii approximation \cite{brazovskii:75:0}. 
Unlike in the earlier phenomenological Landau-Brazovskii
theories \cite{brazovskii:75:0,podneks:96:0}, all parameters in our theory have 
precise relation with measurable quantities. For this reason our density-functional theory allows for 
predicting phase diagrams and EOS in standard thermodynamic variables for given interactions between the particles.

In the case of simple fluids, with dominant attractive interactions between the particles, the MF theories
predict correct topology of the phase diagram. Only details concerning the shape of the coexistence curve close 
to the critical point are incorrect. In contrast, in the self-assembling systems the topology of the MF 
phase diagram is incorrect. Only at low temperature the sequence of ordered phases in MF and in simulations agree.
When temperature increases, the periodic structures loose 
stability one by one, whereas in MF they all are stable up to the same temperature; only the range of density
corresponding to the stability of the ordered phases decreases for increasing $T$.

Let us discuss the physical reason for the qualitatively 
incorrect predictions of the MF theories at relatively high $T$, and the 
physical meaning of the fluctuation-contributions in our theory. In the case of the disordered phase,
the average volume fraction is position-independent, and the MF internal energy is 
$\frac{1}{2}\bar \zeta^2\int d{\bf r}V_{co}(r)$. In a homogeneous structure, i.e. when 
 the particles are more or less homogeneously distributed in space in majority of microstates, 
this is a fair approximation. However, in the case of competing interactions, 
the homogeneous distribution of particles 
occurs only at very high temperature or at very low density. At moderate temperature,
the particles are not homogeneously distributed in the D phase, 
and aggregates are formed in majority of the microstates, as can be seen in
simulation snapshots, cluster analysis~\cite{almarza:14:0,zhuang:16:0,zhuang:16:1,santos:17:0}, 
and in the cartoon in Fig.\ref{cartoon}b.  Thus, in the most probable microstates
the distribution of the particles is 
significantly different from the position-independent average volume fraction. In a typical microstate, 
there are much more particle pairs  at  distances close to the minimum of the interaction potential, 
and much less particle pairs   at  distances corresponding to the repulsion,
than for a homogeneous distribution of the particles (see Fig.\ref{cartoon}b.)
For this reason, the internal energy in the D phase is much lower than predicted in MF. 
On the other hand, the entropy decreases when the aggregates are formed. 

The 
decrease of both, the internal energy and the entropy that is associated with a presence of delocalized aggregates
should be taken into
account in a correction to the MF expression for the grand potential. To see that it is in fact what we do
by adding our fluctuation corrections obtained by formal considerations, 
let us focus on the $\varphi^4$ theory, and 
Eq.(\ref{OmHarval1}).
Using Eqs.(\ref{Gg}), (\ref{Z}), (\ref{C0}) and (\ref{Ck}), we rewrite the first fluctuation-contribution 
in Eq.(\ref{OmHarval1}) in the form
\begin{equation}
 \label{compi}
2a\sqrt{T^*}Z[\zeta]=\frac{2a\sqrt{T^*}}{Z[\zeta]}Z[\zeta]^2=
{\cal G}\Big(\beta \tilde V_{co}(k_0)+A_2[\bar\zeta]+\frac{A_4[\bar\zeta]}{2}{\cal G}\Big),
\end{equation}
and for the D phase we obtain 
\begin{equation}
 \label{OmHarval3}
\beta\Omega(\bar\zeta)/{\cal V}\approx \beta\Omega_{co}(\bar\zeta)/{\cal V}+\beta\tilde V_{co}(k_0)
{\cal G} +a_2(\bar\zeta){\cal G}+
\frac{3a_4(\bar\zeta)}{8}{\cal G}^2.
\end{equation}

Note that Eq.(\ref{OmHarval3}) is similar to the MF grand potential for a weakly ordered phase in the $\varphi^4$ 
theory (Eq.(\ref{MF})
with $M=4$), except that in (\ref{OmHarval3}) ${\cal G}$ plays a role analogous to $\Phi^2/2$ in (\ref{MF}) 
(recall that $\beta\tilde V_{co}(k_0)=-\beta^*$). 
In this mesoscopic theory, $\sqrt{{\cal G}}$ represents a standard deviation of the  local volume fraction 
from the space-averaged value $\bar \zeta$ (see (\ref{calG})). 
Note also that the main difference between the D phase and the weakly ordered phase is the 
fact that the aggregates in the latter phase fluctuate around their average positions, while in the D
phase they move freely. In both cases, the effect on the internal energy depends on the increase of the local 
density in the aggregates and the decrease of the density between them, i.e. on $\sqrt {\cal G}$ or $\Phi$. 
In this approximation, only the most probable density waves, with the period  $2\pi/k_0$,
and the energy decrease proportional to 
$\tilde V_{co}(k_0)$ are taken into account. In  Eq.(\ref{MF}), 
the energy gain in the weakly ordered phases is associated with the MF {\it average deviation from} $\bar \zeta$.
In our 
fluctuation-contribution to the internal energy of the D phase, the energy gain
is associated with the {\it standard deviation 
of the local density from} $\bar \zeta$. The remaining terms in (\ref{MF}) and
(\ref{OmHarval3}) represent the decrease of
entropy in the presence of inhomogeneities - MF average profiles in (\ref{MF}), 
and delocalized aggregates in (\ref{OmHarval3}).
One can see that in our theory the
fluctuation contribution leads to the decrease of the internal energy and entropy, as expected on physical grounds. 
In the ordered phases the average volume fraction profile is smeared because of the fluctuations 
about the average positions, and the fluctuation contributions play a similar role as in the D phase.  

One could consider better approximations for the direct correlation function in (\ref{Ck}) or (\ref{Ck6}). 
 However, since the present approximation 
captures the main physical effect 
of spontaneously appearing inhomogeneities, the high-temperature part of the phase 
diagram is correctly reproduced,
and the functional (\ref{OmHarval2}) is relatively simple, we think that the present
approximation is a good compromise between 
the accuracy and feasibility. 

In the second part of this work we have developed a simplified theory valid for
weakly-ordered phases, i.e. for the high-temperature 
part of the phase diagram. Predictions of this version of the theory agree quite
well with simulation results, except that we
predict lower temperature range of the stability of the lamellar phase, and the
density range in simulations is too
small to verify the 
stability of the IH and IC phases. We therefore could not verify if the IH phase, 
stable up to higher temperatures than the L phase in our 
theory, has the same property in reality. 

The  stability of the inverse phases with periodically distributed voids to 
higher temperatures than in the case of 
phases with periodically distributed clusters is an unexpected result.
In MF, there is no such difference between the stability 
ranges of the H and IH  phases. Our results show that fluctuations are
more destructive for the periodic order of clusters than for 
the periodic order of voids. 

We finally note that the functional (\ref{OmHarval2}) can be applied not only for 
determination of the phase diagram and EOS,
but also to studies of interfaces between different phases and effects of confinement.

\section{Acknowledgments}
I would like to thank Patrick Charbonneau and Yuan Zhuang for very useful additional information,
discussions and comments.
This project has received funding from the European Union's Horizon 2020 research
and innovation programme under the Marie Sk{\l}odowska-Curie grant agreement No 734276
(CONIN). An additional support in the years 2017-2018 has been granted for the CONIN
project by the  Ministry of Science and Higher Education of Poland. Financial support from the
National Science Center under grant No. 2015/19/B/ST3/03122 is also acknowledged.
\section{Appendix}
\subsection{The ordered structures in the one-shell approximation}
The expressions for the first shells of the phases H, IH and C, IC (with the bcc symmetry), Eq.(\ref{g}),
in the real-space representation are~\cite{ciach:10:1}
\begin{eqnarray}
\label{hex}
g^{hex}_1({\bf r})=\sqrt{\frac{2}{3}}\Bigg[\cos(k_br_1)+
2\cos\Big(\frac{k_br_1}{2}\Big)\cos\Big(\frac{\sqrt 3 k_br_2}{2}\Big)\Bigg]
\end{eqnarray}
\begin{eqnarray}
\label{bcc}
g^{bcc}_1({\bf r})=\frac{1}{\sqrt 3}\sum_{i<j}
\Bigg(\cos\Big(\frac{k_b(r_i+r_j)}{\sqrt 2}\Big)
+\cos\Big(\frac{k_b(r_i-r_j)}{\sqrt 2}\Big)\Bigg),
 \end{eqnarray}
where ${\bf r}=(r_1,r_2,r_3)$. 
The geometric factors (\ref{kappa}) for the considered phases are the following:
$\kappa_2=1$ for all the structures, and
\begin{eqnarray}
 {\rm L:}\hskip0.5cm \kappa_3=0, \kappa_4=\frac{3}{2}, \kappa_5=0,\kappa_6=\frac{5}{2}
\end{eqnarray}

\begin{eqnarray}
 {\rm H, IH:}\hskip0.5cm \kappa_3=\sqrt{\frac{2}{3}},\kappa_4=\frac{5}{2}, \kappa_5=5\sqrt{\frac{2}{3}},
 \kappa_6=\frac{85}{9}
\end{eqnarray}

\begin{eqnarray}
 {\rm C, IC:}\hskip0.5cm \kappa_3=\frac{2}{\sqrt{3}},\kappa_4=\frac{15}{4}, \kappa_5=5\sqrt{3},
 \kappa_6=\frac{220}{9}
\end{eqnarray}

\subsection{Explicit expressions for Eqs.(\ref{eqP}), (\ref{eqz}) and for pressure in the case of weak order}
In the $\varphi^M$ theory,   Eqs.(\ref{eqP}) and (\ref{eqz}) take the explicit forms 
\begin{eqnarray}
\label{P}
 a_2(\bar\zeta)-\beta^*+\sum_{n=3}^M\frac{a_n(\bar\zeta)\kappa_n\Phi^{n-2}}{(n-1)!}\\
 \nonumber
 + \frac{a\sqrt T^*}{Z(\bar\zeta,\Phi)}\sum_{n=4}^M\frac{a_n(\bar\zeta)\kappa_{n-2}\Phi^{n-4}}{(n-3)!}
 +\frac{a_6(\bar\zeta)a^2 T^*}{2Z(\bar\zeta,\Phi)^2}=0,
\end{eqnarray}
 and
\begin{eqnarray}
\label{mu}
\mu^*=\mu/|\tilde V(k_0)|=v_0\bar\zeta +T^*\Bigg[a_1(\bar\zeta)+
 \sum_{n=2}^M\frac{a_{n+1}(\bar\zeta)\kappa_n\Phi^n}{n!}\Bigg]\\
 \nonumber 
 +T^*\Bigg[
 \Big(a_3(\bar\zeta)+\sum_{n=2}^{M-2}\frac{a_{n+3}(\bar\zeta)\kappa_n\Phi^n}{n!} \Big)
 \frac{a\sqrt T^*}{Z(\bar\zeta,\Phi)}+
\Big( a_5(\bar\zeta)+\frac{a_7(\bar\zeta)\Phi^2}{2}\Big)
\frac{a^2 T^*}{2 Z(\bar\zeta,\Phi)^2}\Bigg]
\end{eqnarray}
where  in the $\varphi^4$-theory, the terms proportional to $a_n(\bar\zeta)$ with $n>4$ in (\ref{P}), and
$n>5$ in (\ref{mu})
must be disregarded. In obtaining (\ref{P}) and (\ref{mu}), we have  used Eqs. (\ref{MF}),
(\ref{fhn}), (\ref{calG}) and  (\ref{Gg}).
$Z(\bar\zeta,\Phi)$ is given by (\ref{sqrtr}) and (\ref{W}), or by (\ref{Y}) in the $\varphi^4$ or
$\varphi^6$ theory respectively,
 with $A_n[\zeta]$ 
approximated by  (\ref{An}). 
The thermodynamic pressure $p=-\Omega/V$ is given by the equation
(see (\ref{OmHarval1}) or (\ref{OmHarval2}), and (\ref{MF}))
\begin{eqnarray}
 \label{pres}
 p^*=p/|\tilde V(k_0)|=-\frac{v_0\bar\zeta^2}{2}+\mu^*\bar\zeta-T^*\Bigg[
 \beta f_h(\bar\zeta)+\frac{a_2(\bar\zeta)-\beta^*}{2}\Phi^2
 +\sum_{n=3}^M\frac{a_n(\bar\zeta)}{n!}\kappa_n\Phi^n\\
 \nonumber
 + 2a\sqrt T^*Z(\bar\zeta,\Phi)
 -\frac{a^2T^*}{2Z(\bar\zeta,\Phi)^2}\Big(a_4(\bar\zeta)+\frac{a_6(\bar\zeta)\Phi^2}{2} \Big)
 -\frac{a_6(\bar\zeta)a^3T^{*3/2}}{3Z(\bar\zeta,\Phi)^3}
 \Bigg],
\end{eqnarray}
where $\Phi$ satisfies (\ref{P}) and $\mu^*$ is given in (\ref{mu}). In the $\varphi^4$-theory, 
the terms proportional to $a_n(\bar\zeta)$ with $n>4$
must be disregarded in (\ref{pres}).

\end{document}